\documentclass[a4paper,fleqn]{cas-dc}

\usepackage[numbers]{natbib}
\usepackage[normalem]{ulem}

\def\tsc#1{\csdef{#1}{\textsc{\lowercase{#1}}\xspace}}
\tsc{ARPES}
\newcommand{\PbSnSe}{Pb$_{1-x}$Sn$_x$Se}

\begin{document}
\let\WriteBookmarks\relax
\def\floatpagepagefraction{1}
\def\textpagefraction{.001}

\shorttitle{}
\shortauthors{B. Turowski et~al.}

\title [mode = title]{Spin-polarization of topological crystalline and normal insulator \PbSnSe~(111) epilayers probed by photoelectron spectroscopy}                      
\tnotemark[1]                    

\author[1]{Bartłomiej Turowski}[orcid=0000-0003-1229-4416]
\credit{Writing - original draft, Investigation, Analysis}
\author[1]{Aleksandr Kazakov}[orcid=0000-0002-8066-2922]
\credit{Investigation, Analysis, Writing - review \& editing}
\author[1]{Rafał Rudniewski}[]
\credit{Investigation}
\author[2]{Tomasz Sobol}[]
\credit{Investigation}
\author[2]{Ewa Partyka-Jankowska}[]
\credit{Investigation}
\author[1]{Tomasz Wojciechowski}[orcid=0000-0002-6424-988X]
\credit{Investigation}
\author[3]{Marta Aleszkiewicz}[orcid=0000-0002-4939-9855]
\credit{Investigation}
\author[1]{Wojciech Zaleszczyk}[orcid=0000-0002-0051-7267]
\credit{Investigation}
\author[2]{Magdalena Szczepanik}[]
\credit{Investigation, Supervision}
\author[1]{Tomasz Wojtowicz}[orcid=0000-0003-4498-4622]
\credit{Supervision, Funding acquisition, Writing - review \& editing}
\author[1,4]{Valentine V. Volobuev}[orcid=0000-0002-8474-7392]
\credit{Writing - original draft, review \& editing, Conceptualization, Investigation, Analysis, Supervision}
\cormark[1]
\ead{volobuiev@magtop.ifpan.edu.pl}

\address[1]{International Research Centre MagTop, Institute of Physics, Polish Academy of Sciences, Aleja Lotników 32/46, Warsaw, PL-02668, Poland}
\address[2]{National Synchrotron Radiation Centre SOLARIS, Jagiellonian University, Czerwone Maki 98, Kraków, PL-30392, Poland}
\address[3]{Institute of Physics, Polish Academy of Sciences, Aleja Lotników 32/46, Warsaw, PL-02668, Poland}
\address[4]{National Technical University "KhPI", Kyrpychova Str. 2, Kharkiv, 61002, Ukraine}

\cortext[cor1]{Corresponding author at: International Research Centre MagTop, Institute of Physics, Polish Academy of Sciences, Aleja Lotników 32/46, PL-02668 Warsaw, Poland}

\begin{abstract}
The helical spin texture on the surface of topological crystalline insulators (TCI) makes these materials attractive for application in spintronics. In this work, spin-polarization and electronic structure of surface states of (111)-oriented \PbSnSe\ TCI epitaxial films are examined by angle- as well as spin-resolved photoemission spectroscopy (SR-ARPES). High-quality epilayers with various Sn content are grown by the molecular beam epitaxy (MBE) method. Topological - normal insulator transition manifesting itself as band gap opening is observed. It is shown that the gap opening can be induced not only by changing the Sn content of the epilayer but also depositing a transition metal (TM) on its surface. In the latter case, the observed gaping of the surface states is caused by change in surface composition and not by magnetism. We also show that helical spin polarization is present not only for samples of topological composition but also for trivial ones (with an open band gap). The observed spin polarization reaches a value of 30\% for the in-plane spin component and is almost absent for the out-of-plane one. We believe that our work will pave the way for the application of surface states not only of topological but also normal insulators based on lead-tin chalcogenides in spin-charge conversion devices.
\end{abstract}


\ExplSyntaxOn
\keys_set:nn { stm / mktitle } { nologo }
\ExplSyntaxOff
\maketitle
\begin{sloppypar}

\section{Introduction}
With the remarkable discovery of topological insulators (TIs) the new era of materials has begun, the era of topological materials \cite{bernevig2006quantum, konig2007quantum}. Recently, many new topologically protected phases of matter have been discovered in various dimensions from 1D to 3D. These phases include Dirac, Weyl, nodal and chiral semimetals, higher order TIs, etc. \cite{wieder2022topological, narang2021topology}. Currently, scientists are actively seeking  practical applications using topologically protected states to boost device performance \cite{nayak2008non,dey2021recent,smirnova2020nonlinear, wang2020recent, li2020heterogeneous, yang2020topological, ZHANG2020145290, skinner2018large}. For instance, one of the possible application of TIs is spintronics \cite{liu2014spin, he2022topological}. Due to their spin-momentum locked helical nature, the surface states of 3D TIs can be used as efficient spin-charge converters \cite{Li_nature_nano2014, wang2016surface, dey2021recent}. Moreover, as a proof of concept, controllable magnetization switching by spin-orbit torque induced by spin polarized TSS has been successfully demonstrated in TI/ferromagnet heterostructures \cite{mellnik2014spin, ding2021switching, han2021topological}. Therefore, knowledge of spin-structure of topological materials is of great importance for the development of the field of spintronics.

Despite its amazing properties, the spin texture of many interesting topological materials remains relatively unexplored. Especially fascinating for spin-resolved experiments is the topological - normal insulator transition, where the spin texture of topological material is expected to evolve. It has been proved experimentally that helical spin texture can be preserved even for the normal insulator phase in these materials \cite{Xu2015, wojek2013spin}. This fact definitely widens the range of the materials suitable for spintronic device applications since it does not require band inversion. Topological crystalline insulators (TCIs) \cite{fu2011topological}, in which topological protection originates from crystal symmetry, are very sensitive to external perturbations unlike those Z$_2$-TIs protected by the time-reversal symmetry \cite{ando2015topological}. Therefore, it is easy to observe various topological phases \cite{liang2017pressure,lau2019topological, schindler2018higher, mandal2017topological}, as well as TCI - normal insulator (NI) transition \cite{Dziawa2012, mandal2017topological, volobuev2017giant, assaf2017magnetooptical, phuphachong2017dirac, kazakov2021signatures, assaf2022acsphotonics} induced in these materials by various means, e.g. manipulating their crystal structure and lattice parameter. TCIs, similarly to TIs, possess topological surface states (TSS) and helical spin-momentum locked texture \cite{wojek2013spin, xu2012observation, safaei2013topological}. Prototypical TCI based on Sn monochalcogenide have a rock-salt crystal structure with TSS protected by mirror symmetry of \{110\} planes \cite{hsieh2012topological}. The band ordering can be tuned from inverted into a trivial one by alloying TCI with NI material, e.g. Pb monochalcogenide \cite{assaf2017magnetooptical,Dziawa2012,kazakov2021signatures,mandal2017topological, volobuev2017giant,galeeva2021photoelectromagnetic}. Alternatively, since the value of the band gap is strongly temperature dependent, band inversion can be easily removed by heating of lead-tin chalcogenide alloy. Topologically protected SS exist on \{100\}, \{110\} and \{111\} faces of TCI crystals \cite{ando2015topological, hsieh2012topological}. \{100\} TSS can be easily observed by ARPES because it is the natural cleavage plane of TCI crystals \cite{Dziawa2012, xu2012observation}. To obtain the \{111\} surface, growth by MBE is often exploited \cite{assaf2017magnetooptical, kazakov2021signatures, Krizman2018, mandal2017topological, volobuev2017giant}. In the latter case topologically protected Dirac cones are revealed in the vicinity of the $\overline{\Gamma}$ and  $\overline{\mathrm{M}}$ points \cite{ando2015topological, mandal2017topological}. The helical spin structure similar to prototypical 3D TI Bi$_{2}$Se$_{3}$ is expected in (111)-oriented \PbSnSe\ TCI \cite{safaei2013topological}.

Among TCIs, \PbSnSe\ mixed crystals are of significance for applications due to their low carrier density and relatively high mobility \cite{assaf2017magnetooptical, phuphachong2017dirac}. The carrier type can be easily controlled and n-type samples required for conventional ARPES studies to observe states of both conduction (CB) and valence band (VB) can be grown by MBE \cite{mandal2017topological, polley2014observation}. In this work, spin-polarized band structure of (111) \PbSnSe\ is studied by SR-ARPES across the topological transition. To our knowledge there are only several works where the spin texture was studied in both TI and NI states \cite{Xu2015, wojek2013spin, sanchez2018anomalous}.  We expect that our work will stimulate further research toward understanding the spin texture of TCIs and exploring the possibility of using lead-tin chalcogenide based TCIs as active elements of spintronic devices.

\section{Experimental methods}

\subsection{Epilayers preparation}

\begin{figure} [h]
	\centering
		\includegraphics[scale=0.93]{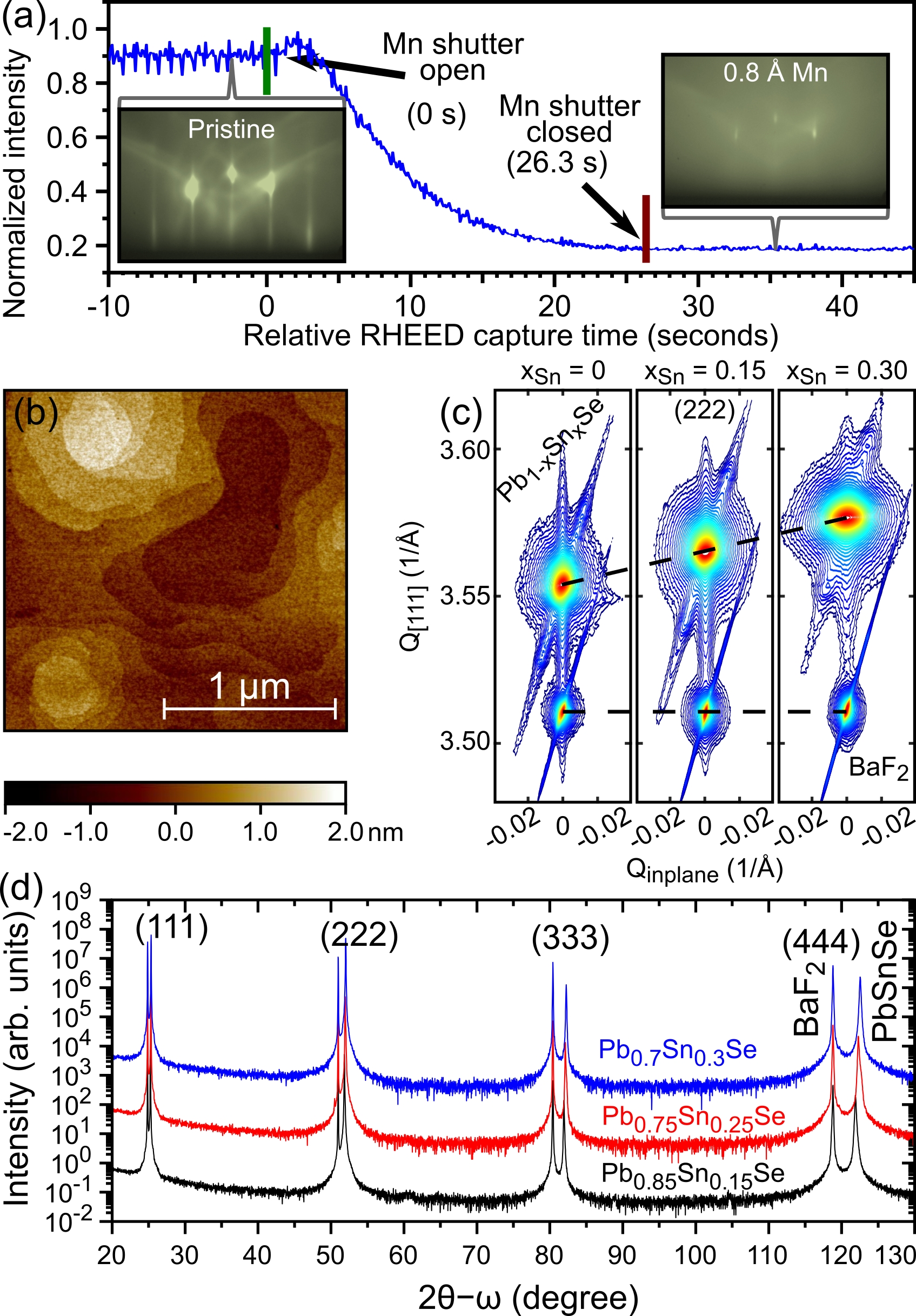}
	\caption{ Structural investigation of (111) TCI epilayers. (a) Specular spot intensity as a function of Mn deposition. Left inset shows RHEED pattern of 1 \textmu m Pb$_{0.7}$Sn$_{0.3}$Se epilayer recorded along [110] azimuths, right inset is RHEED pattern of the same sample after 0.8~\AA\ of Mn deposition, (b) AFM image of the same sample, (c) XRD reciprocal space maps (RSM) of symmetric (222) reflection obtained for \PbSnSe\ with Sn content of 0\%, 15\% and 30\%, respectively left to right, (d) XRD 2$\theta-\omega$ scans for the 3 samples with the same compositions as the samples used in the ARPES measurements.}
	\label{FIG:1}
\end{figure}

The n-type \PbSnSe\ 1~\textmu m thick epilayers were grown by MBE in Veeco GENxplor growth chamber, with base pressure bellow 10$^{-10}$~mbar, using effusion cells with stoichiometric PbSe, SnSe as sources. Using Bi$_{2}$Se$_{3}$ effusion cell, bismuth doping  in the range 0.07-0.1 at. \% was employed to tune Fermi energy and obtain n-type films. Half millimeter thick (111)-oriented BaF$_{2}$ plates, freshly cleaved from the bulk crystal of Korth Kristalle GmbH, were used as a substrate. Epitaxial growth of \PbSnSe\ with x$_{Sn}$ = 0.15, 0.25 and 0.30 was conducted at 350 \textdegree C, with material flux ratio and deposition rate controlled by quartz crystal microbalance (QCM) placed in front of the substrate position. The deposition rate was determined to be 0.2~nm/sec. Real-time \textit{in-situ} monitoring of structural quality was done by reflection high-energy electron diffraction (RHEED) (see Fig.~\ref{FIG:1} (a)). After the growth process, without breaking ultra-high vacuum (UHV), samples were transferred to Ferrovac VSN40S UHV suitcase (with a base pressure of 2$\times$10$^{-11}$~mbar) for transport to the synchrotron facility \cite{volobuev2017giant,mandal2017topological, berchenko2018surface, tarasov2021surface}. The suitcase was later attached to the preparation chamber of SOLARIS PHELIX beamline end-station \cite{SZCZEPANIKCIBA202149} to transfer the samples to the measurements setup without breaking the vacuum.

\subsection{Characterization techniques}

\begin{figure} [h]
	\centering
		\includegraphics[scale=0.93]{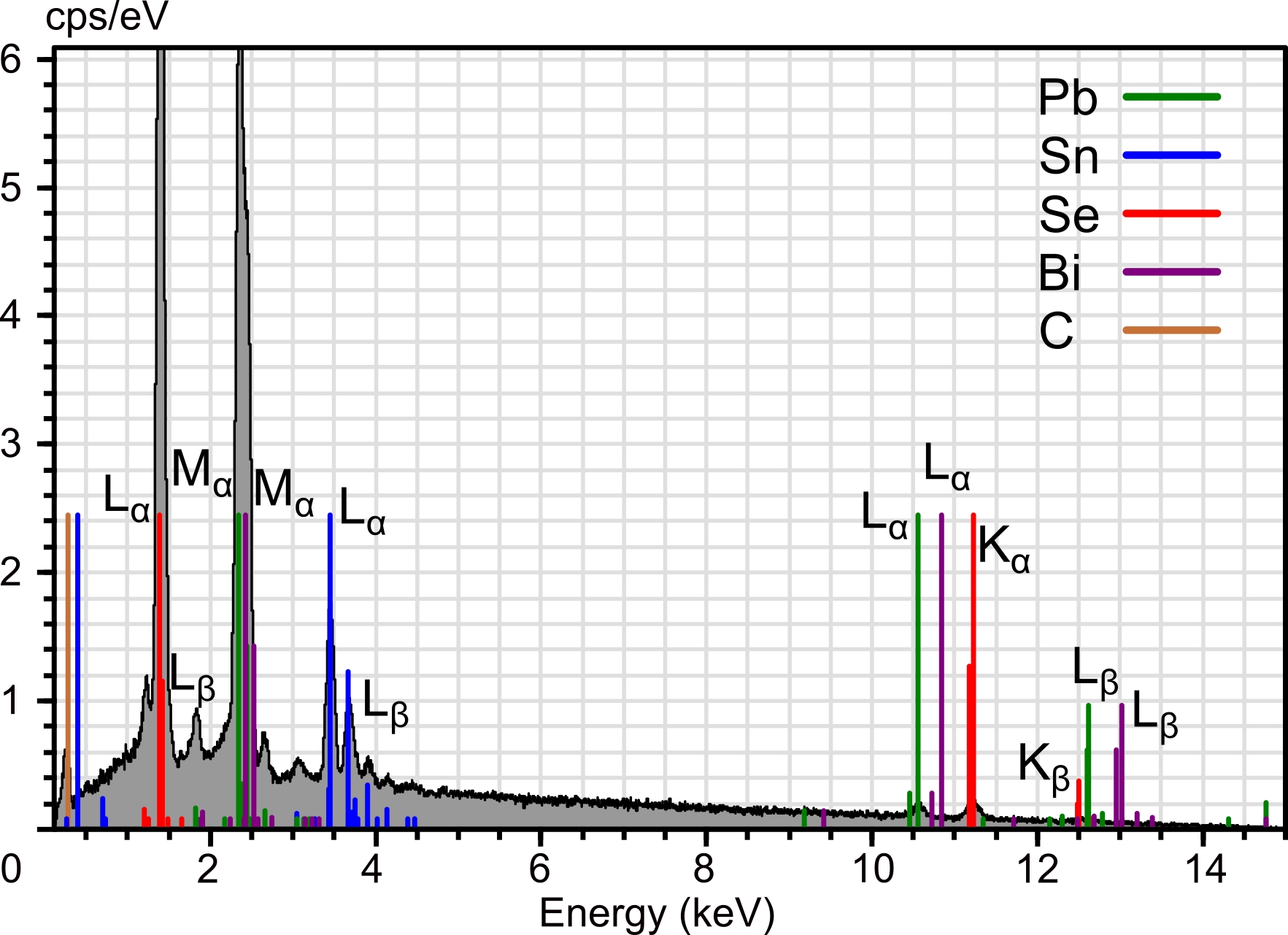}
	\caption{EDX of 1 \textmu m {\PbSnSe} film with x$_{Sn}$ = 0.32.
	Operating voltage 15 keV.}
	\label{FIG:2}
\end{figure}

Atomic force microscopy (AFM) images (Fig.~\ref{FIG:1} (b))  were obtained in tapping mode with Bruker MultiMode 8-HR microscope. Composition and additional morphology characterization were done with Zeiss Auriga Cross-Beam Neon 40 Scanning electron microscope (SEM) equipped with QUAN-TAX 400 Bruker Electron Dispersive X-ray spectroscopy (EDX) system (Fig.~\ref{FIG:2}). X-ray diffraction (XRD) patterns (Fig.~\ref{FIG:1} (c)) were recorded by PANalytical X'Pert Pro MRD diffractometer with a 1.6~kW x-ray tube using CuK$\alpha _{1}$ radiation ($\lambda$ = 1.5406~\AA), a symmetric 2 $\times$ Ge (220) monochromator and 2D Pixel detector. 

The electronic band structure and spin polarization of the films were examined by ARPES and SR-ARPES at PHELIX Beamline of National Synchrotron Radiation Centre SOLARIS in Krakow (Poland). The end-station is equipped with SPECS GmbH PHOIBOS 225 electrostatic hemi-spherical analyzer and an automated six-axis cryogenic manipulator. Angular and energy resolutions of the spectrometer were 2~meV and 0.1~\textdegree, respectively, with the excitation spot diameter of hundred \textmu ms. For SR-ARPES measurements FOCUS GmbH FERRUM 2D VLEED (Very Low Energy Electron Diffraction) spin detector \cite{escher2011ferrum}, with an additional spin rotator installed before the entrance flange giving access to all three spin components, was used. Detailed information about this end-station may be found in Ref. \cite{SZCZEPANIKCIBA202149}. The studied samples were probed by ARPES at the temperature range of 78-298~K with a variable photon energy of 50-90~eV. The intensity of spin-dependent energy profiles was recorded at least at two opposite values of \textit{k} at 200~K with 70~eV photon energy. Both ARPES and SR-ARPES measurements were performed with p-polarized light in the vicinity of the $\overline{\Gamma}$ point in \textit{$\overline{\Gamma}$} -  $\overline{\mathrm{M}}$ direction.

To obtain the values of the spin polarization $P$, the asymmetry of the spin-polarized electrons $A$ was first calculated using \cite{okuda2017recent}:
\begin{equation}
A = \frac{I_{+}-I_{-}}{I_{+}+I_{-}},
\end{equation}
where $I_{+}$ and $I_{-}$ are intensities of energy dispersion curves (EDCs) for VLEED target magnetized in opposite directions. Then polarization can be derived as:
\begin{equation}
P = A/S,
\end{equation}
where $S=0.275$ is Sherman function coefficient \cite{Sherman1956}, determined for this detector by measuring a fully polarized electron beam. Finally, spin-resolved EDCs are obtained as follows \cite{osterwalder2006spin, bigi2017very}:
\begin{equation}
SEDC_{\uparrow,\downarrow} = \frac{I_{+}+I_{-}}{2}\cdot (1\pm P).
\end{equation}

\subsection{Transition metal deposition on the surface of the epilayers}

Uncompensated magnetic moments of transition metals (TM) on a surface of TCI may break the mirror or time-reversal symmetry and open band gap as was predicted by theory \cite{fang2014largeQAHE, serbyn2014symmetry, reja2017surfaceM, fertig2019probingM, kazakov2021signatures}. To test these predictions, as well as to verify the effect of TM surface doping on the TCI spin texture, iron (Fe) and manganese (Mn) were deposited on the surfaces of three epilayers in amounts given in Table~\ref{tbl1}. In both cases, the thickness of deposited TM layers was monitored by calibrated QCM. The condensation rate was measured just before TM deposition on the sample surface. Mn deposition was carried out in the growth chamber of the Veeco GENxplor system at room temperature (RT) to achieve less than 0.5 monolayer (ML) coverage using Knudsen effusion cell and the samples were delivered to the synchrotron in the UHV suitcase. Fe depositions were performed at the synchrotron facilities, in the preparation chamber of the PHELIX beamline end-station using an available electron beam evaporator to obtain submonolayer surface coverage. After each Fe deposition, samples were immediately transferred to the analytical chamber, where ARPES E(k) spectra and SR-ARPES EDCs were recorded.

\begin{table}[width=.9\linewidth,cols=3,pos=h]
\caption{Deposition  of transition metals (thickness in {\AA}) on surface of \PbSnSe\ epilayers}\label{tbl1}
\begin{tabular*}{\tblwidth}{LLLLL@{} LLLLL@{} }
\toprule
Sample & TM & 1$^{st}$ dep. & 2$^{nd}$ dep. & Sum\\
\midrule
A - 25\% Sn & Mn & 0.3 & - & 0.3 \\
B - 25\% Sn & Fe & 0.2 & - & 0.2 \\
C - 30\% Sn & Fe & 0.15 & 0.3 & 0.45 \\
\bottomrule
\end{tabular*}
\end{table}

\section{Results and discussion}
\subsection{Structural and compositional characterization}

{\PbSnSe} epilayers that were used in current ARPES and SR-ARPES investigations had high quality similar to that of structures used in previously reported works \cite{assaf2017magnetooptical, mandal2017topological, Krizman2018, kazakov2021signatures, phuphachong2017dirac}. Streaky and high contrast RHEED patterns with reflections on Laue semicircle and intense Kikuchi lines (Fig. \ref{FIG:1} (a)) point to smooth two-dimensional (2D) growth of the films that result in an atomically flat surface. It is consistent with clearly defined contours of atomically flat terraces visible in the AFM image (Fig. \ref{FIG:1} (b)). The average RMS roughness is estimated to be below 1 nm. The growth occurs in a 2D step-flow mode (see previous growth reports \cite{Springholz1996, wu2006scanning}). The lattice mismatch between \PbSnSe\ and BaF${_2}$  in a range of 1.2-2\% causes screw type threading dislocations, at which the surface steps are pinned. These dislocations lead to the formation of a characteristic spiral step structure visible in Figure \ref{FIG:1} (b). XRD further confirms the growth of single-phase single-crystalline films. For three samples of the composition used in this work for ARPES measurements, only \{111\}-type of reflections of the substrate and the film are present (Fig.~\ref{FIG:1}~(d)). The typical FWHM value of $\omega$-rocking curve extracted from (222) RSM (Fig.~\ref{FIG:1}~(c)) is determined to be 200-500 arcsec for 1 \textmu m thick \PbSnSe\ films while it equals to 40-70 arcsec for substrates. At the same time, the position of the of \PbSnSe\ (see the same figure) is a linear function of Sn content. Thus, Sn concentration ($x_{Sn}$) in obtained samples is determined by XRD \cite{Krizman2018}:
\begin{equation}
 x_{Sn} = \frac{6.1240 - a_{PbSnSe} } {0.1246},
\end{equation}
where $a_{PbSnSe}$ is \PbSnSe\ lattice constant. In addition, elemental EDX analysis is used to characterise the composition of samples. It shows the presence of only Pb, Sn and Se in the films (Figure \ref{FIG:2}). However, the amount of Bi below 1 at. \% is beyond detection range of EDX analysis. Small traces of C that are usually present for every sample loaded in SEM are also detected. The composition of the samples estimated from deposition rate measurement agrees within 2\% to the composition obtained from EDX and XRD measurements. Upon TM deposition, we found an increase of background noise (see right inset in Fig.~\ref{FIG:1}(a)) as also manifested by a decrease of intensity of the specular spot on the RHEED pattern (Fig.~\ref{FIG:1}(a)). No additional spots or rings were observed by RHEED which would indicate the formation of disordered (nanocrystalline or amorphous) TM adsorbate on \PbSnSe\ surface.

\subsection{Spin polarization in pristine TCI and NI epilayers}

\begin{figure*}
	\centering
		\includegraphics[scale=0.93]{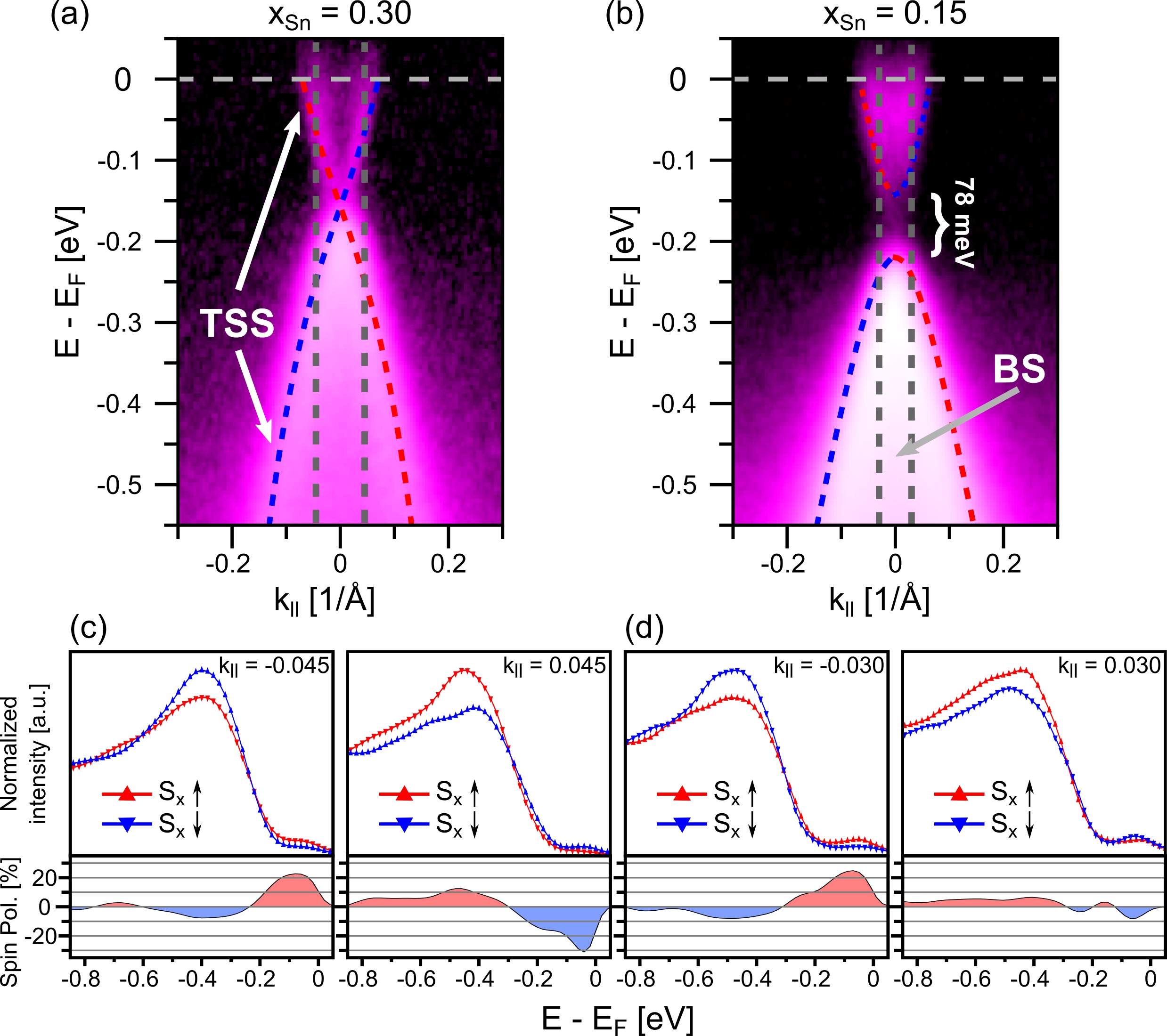}
	\caption{(a), (b) ARPES spectra of TCI \PbSnSe\ (111) epilayers and (c), (d) corresponding SR measurements taken at 200~K with a photon energy of 70~eV in the vicinity of the $\overline{\Gamma}$ point. Color is scaled logarithmically on the images. (a) corresponds to topological x$_{Sn}$ = 0.30 composition with gapless topological surface states (TSS) and (b) to trivial  x$_{Sn}$ = 0.15 composition with band gap of 78 meV and high intensity of bulk states (BS). Red and blue dashed lines are guidelines of the surface states, whereas grey vertical dashed lines indicate negative and positive momentum positions where SR measurements are taken. (c), (d) corresponding to (a) and (b) SEDC of spin-up (red) and spin-down (blue) in-plane spin component and spin polarization is depicted at the bottom. }
	\label{FIG:3}
\end{figure*}

Electronic structure of two \PbSnSe\ samples containing 30\% and 15\% of Sn near $\overline{\Gamma}$ point of surface BZ at 200~K is presented in Fig. \ref{FIG:3} (a) and (b). At this temperature, the studied system turns out to be in the topological (x$_{Sn}$ = 0.30) or normal insulator phase(x$_{Sn}$ = 0.15) \cite{Dziawa2012, Preier1979, Krizman2018} depending on tin content. The TSS crossing at Dirac Point (DP) is visible in the presented TCI spectrum (Fig.\ref{FIG:3} (a)).

The observed band gap of 78~meV in trivial phase (Fig.\ref{FIG:3} (b)) is in agreement with the semi-empirical Preier's formula \cite{Preier1979}. The SR measurements are performed for three orthogonal spin components. Typically in topological insulators helical in-plane spin-polarization is expected. Therefore, the spin is supposed to be tangential to the Fermi surface and consequently the out-of-plane and one in-plane component should to be equal to 0. The corresponding SEDCs are extracted at momentum positions marked by vertical dashed grey lines in Fig. \ref{FIG:3} (a, b). We found negligible spin polarization in the direction normal to the sample plane as well as for a one in-plane component as anticipated. The obtained results for the other in-plane spin components, $S_x$, are presented in Fig. \ref{FIG:3} (c) and (d). As expected, it shows helical spin texture, i.e. opposite spin polarization $P$ for the opposite wave vectors $k_{||}$ for the TCI case (Fig.\ref{FIG:3} (c)). In CB, the spin polarization of TSSs reaches moderate values, up to 30\%. VB TSSs have opposite helicity with respect to the CB, however the value of spin polarization in the VB drops to 10\%. Such reduced values are due to the overlap of VB TSS with spin-degenerate bulk states. Previously in sister material, in (100) oriented PbSnTe \cite{xu2012observation} same spin polarization of 10\% was reported for one of the in-plane components in VB. Typically, the values of spin-polarization is reduced to \textasciitilde50\% due to spin-orbit coupling (SOC) for prototypical Bi$_{2}$Se$_{3}$ and Bi$_{2}$Te$_{3}$ Z2 TIs as shown by DFT \cite{YazevSpin}. Strong SOC is certainly present in \PbSnSe\ and responsible for band inversion in these materials. Therefore a similar effect of reduction of measured spin-polarization is expected.

\begin{figure*}
	\centering
		\includegraphics[scale=0.93]{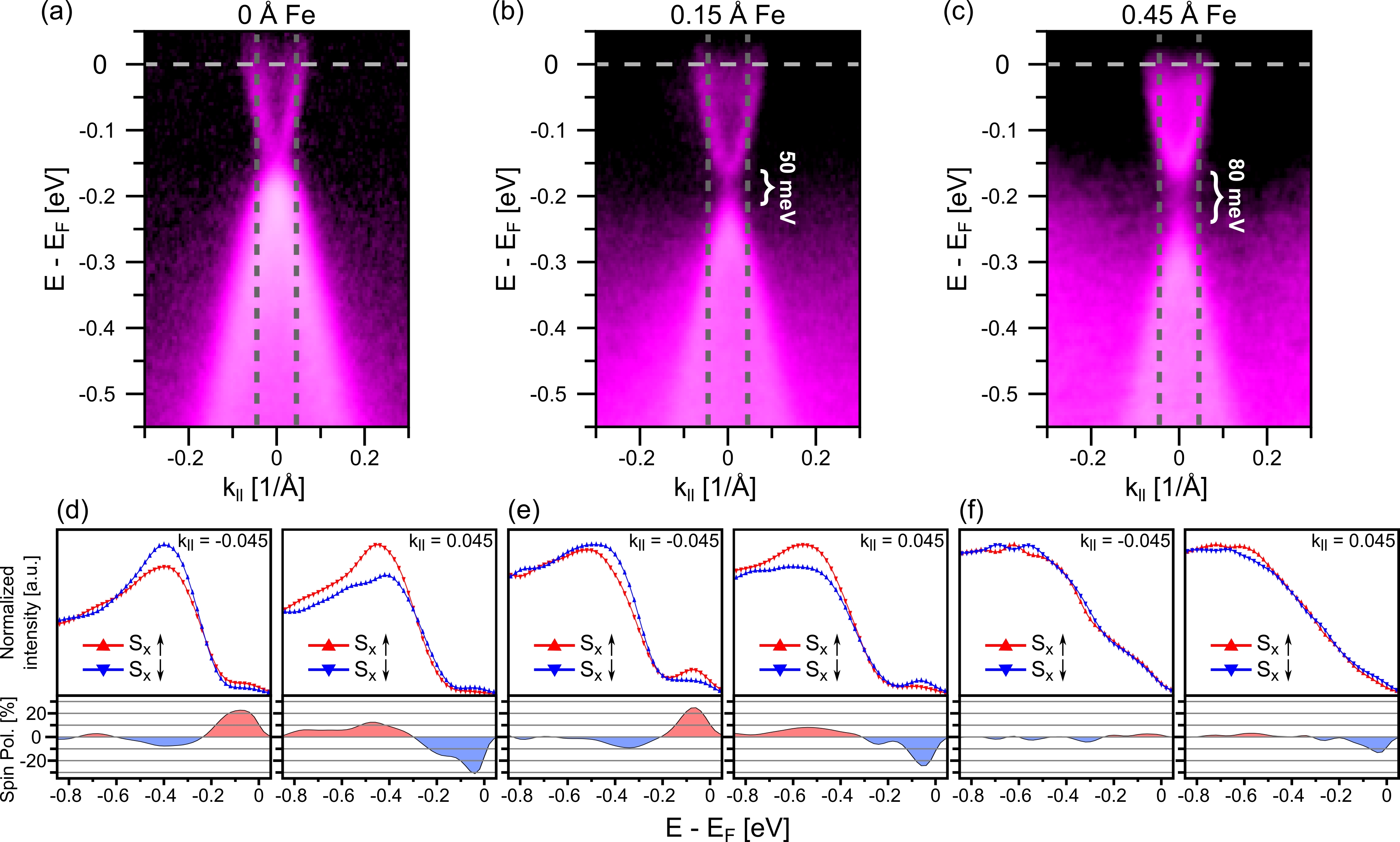}
	\caption{(a)-(c) Electronic structure of Pb$_{0.70}$Sn$_{0.30}$Se taken at the same conditions as the one presented in Fig. \ref{FIG:3} with increasing iron thickness from 0 to 0.15~\AA\ and then to 0.45~\AA\ deposited on epilayer's surface. Color is scaled logarithmically on the images; for (c) strong background noise is removed. Gray vertical dashed lines indicate negative and positive positions of momentum for SR measurements. (d)-(f) Corresponding SEDCs and spin polarization at |k$_{||}|$ = 0.045~1/\AA.}
	\label{FIG:4}
\end{figure*}

Helical spin polarization of the same magnitude is observed for the trivial Pb$_{0.85}$Sn$_{0.15}$Se sample in which the band gap is open (Fig.\ref{FIG:3} (d)). This remarkable effect is consistent with  previous reports for the material systems in which TI-NI transition was observed \cite{Xu2015, wojek2013spin}. Previously, such an effect explained by the presence of Rashba-split surface states that also have helical spin texture. The effect was modelled for Bi$_2$Se$_3$ TI through tight binding calculations. In addition to the TI hamiltonian, Rashba term and electrostatic surface potential were taken into account \cite{jozwiak2016spin}. As the result of changing the sign of a mass parameter, trivial spin-polarized surface resonances having the same helicity as in TI were obtained for the NI phase. Indeed, the giant Rashba effect has been observed previously in (111)-oriented PbSnTe(Se) TCIs and may be responsible for the observed spin polarization in the NI phase \cite{volobuev2017giant, rechcinski2021structure, ito2020observation}. The prerequisites of the Rashba effect are broken inversion symmetry and the presence of an electric field gradient. These conditions are automatically fulfilled due to polar nature of the (111) surface. However, during our measurements, we did not detect Rashba split states even after changing the energy of incident photons. In addition, no Rashba effect was reported for the  extensively studied in the past (001)-oriented TCIs, where spin-polarized SSs were observed for the NI phase \cite{wojek2013spin}. Apparently, the splitting can be so small that it cannot be resolved by ARPES, though detection of the splitting with spin is still possible. It should be noted that the electrostatic band bending effect that is observed in PbSnTe(Se) TCIs with (111) orientation \cite{volobuev2017giant, rechcinski2021structure} can  itself be the origin of the spin-polarized surface resonances even at zero value of Rashba coefficient as shown in ref.~\cite{jozwiak2016spin}.

Furthermore, observation of helical spin texture in trivial samples is consistent with recent transport measurements and theoretical calculations performed for thin \PbSnSe\ epilayers \cite{kazakov2021signatures}. It is known, that the spin-momentum locking in TI TSSs introduces an additional $\pi$ Berry phase shift between the pair of counter-propagating charge contours at the Fermi surface. It is demonstrated in Ref.~\cite{kazakov2021signatures}, that if inversion symmetry is broken (that naturally occurs on the sample surface) the Berry phase quantized to $\pi$ is attained even for NI sample composition. Weak-antilocalization measurements confirmed presence of the $\pi$ Berry phase for both topological and NI composition of \PbSnSe\ epilayers. It was shown, that symmetries present in the system, rather than topological order, are responsible for the Berry phase effects \cite{kazakov2021signatures}.

\subsection{Effect of transition metal deposition on electronic and spin structure of TCI}

A similar effect of TCI-NI insulator transition is detected when an ultrathin amount of transition metal is deposited on the surface of TCI. Fig. \ref{FIG:4} presents evolution of  electronic structure and the spin texture of x$_{Sn}$ = 0.30 epilayer at 200~K upon deposition of Fe.
Initially, the system is in a topological state, with the band gap closed by TSSs (Fig. \ref{FIG:4} (a)). Deposition of 0.15~\AA\ (Fig. \ref{FIG:4} (b)) causes band bending and introduces additional n-type doping to the sample surface. The Fermi level moves upward and a surprisingly large 50 meV gap opens in the middle of the spectrum. Further deposition of 0.3~\AA\ of Fe (0.45~\AA\ in total) leads to a further slight increase of Fermi energy and wider opening of the surface band gap to 80~meV as presented in Fig. \ref{FIG:4} (c). Despite the observed band gap, the surface states (SSs) or so-called precursor states \cite{Xu2015, sanchez2018anomalous} are still detectable in the ARPES spectra. The surface nature of these states is evidenced by the lack of their dispersion when the energy of the incident photon is varied (not shown). It should be noted that the background noise was subtracted in Fig. \ref{FIG:4} (c) to properly visualise SS and band gap. The pronounced increase of the background noise with Fe deposition suggests a disordered TM state on the surface of the film under study. Fig. \ref{FIG:4} (d), (e), (f) show SR measurements for different amount of deposited Fe. After the first 0.15~\AA\ Fe deposition, despite the gap opening, the helicity of the surface states remains unchanged and large spin polarization up to 25~\% is still detected. Further increasing the ultrasmall amount of deposited Fe to 0.45~\AA\ leads to a drastic decrease in spin polarization to $\sim$5\%, which we link to the presence of unpolarized disordered Fe  states. These disordered states overlap with the SS causing an increase in background noise and consequent a decrease in the spin-polarised spectral weight brought by the precursor states.

\begin{figure}[th!]
	\centering
		\includegraphics[scale=0.93]{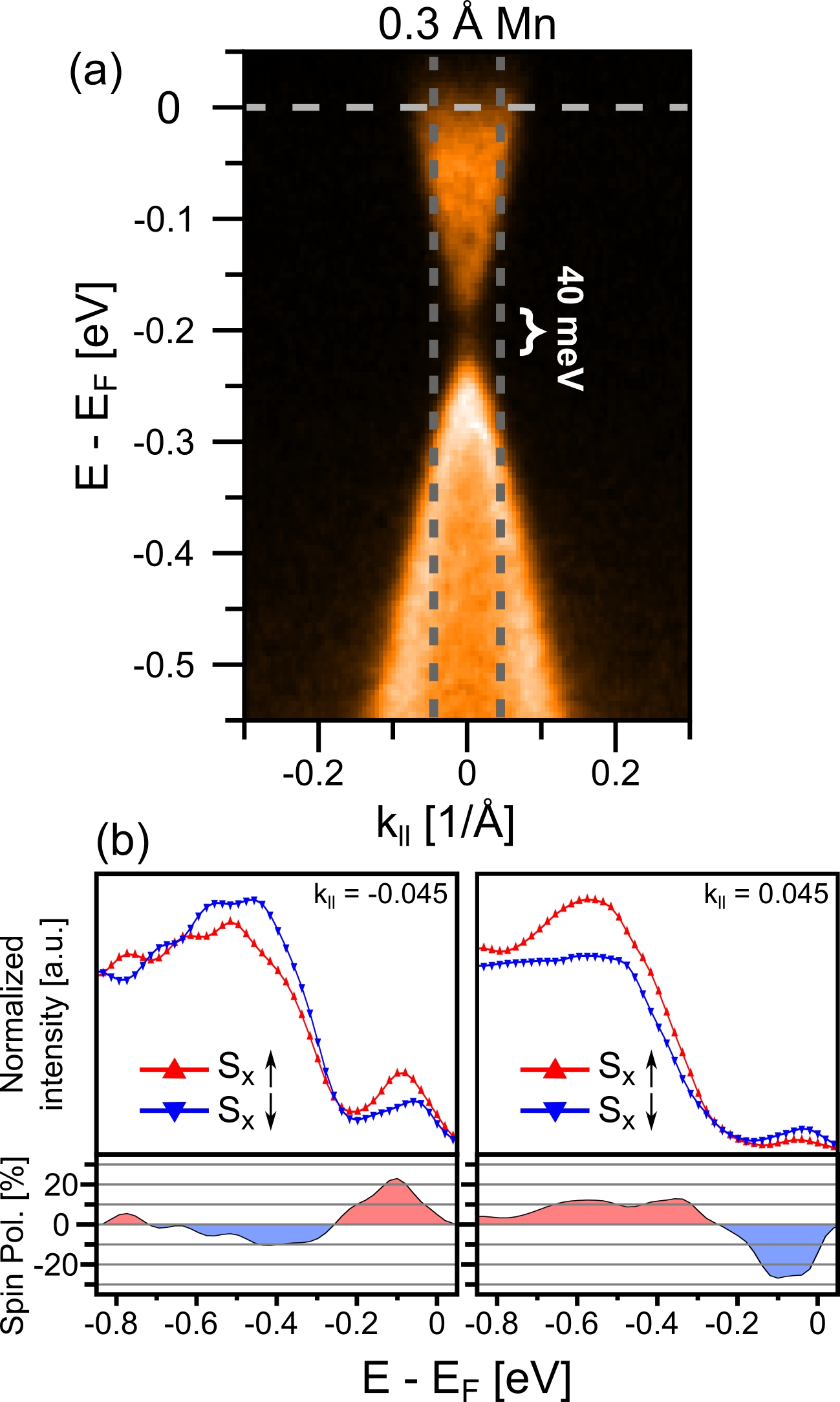}
	\caption{(a) Electronic structure of Pb$_{0.75}$Sn$_{0.25}$Se and (b) corresponding SR measurements taken at 200~K with a photon energy of 70 eV in the vicinity of the $\overline{\Gamma}$ point with 0.3~\AA\ of Mn deposited on the surface. Grey vertical dashed lines indicate the negative and positive momentum of SR measurements. The spectrum are shown with a linear color scale.}	\label{FIG:5}
\end{figure}

The observed gap opening with TM deposition on the TCI surface have been further corroborated by ARPES and SR-ARPES measurements of several samples with topological composition. The spin-polarized SSs, as well as an induced gap, were detected in all cases of deposition of moderate amount of Fe. In order to further confirm the observed effect, we replaced Fe with another TM. Deposition of 0.3~\AA\ of Mn on the surface of Pb$_{0.75}$Sn$_{0.25}$Se yields similar results. Opening of 40~meV SS band gap is observed, as well as spin polarization of up to 30\% (see Figure \ref{FIG:5} (a) and (b) respectively) is detected.

\subsection{The origin of gaped states induced by TM deposition.}

Magnetic impurity on the surface of TCI may cause not only gap opening due to the breaking of time reversal or mirror symmetry \cite{fang2014largeQAHE, serbyn2014symmetry, kazakov2021signatures} but also can create a 2D ferromagnetic state \cite{reja2017surfaceM, fertig2019probingM}. However, in the latter case, the Fermi energy should lie in the gap, not observed in the experiment. Also, there should be a critical temperature of magnetic transition if the gap has a magnetic origin and is related to the arrangement of magnetic moments. Provided magnetic arrangement, below this characteristic temperature the gap should be always open. On the other hand, other mechanisms can cause the band gap opening, e.g. a simple change of band inversion point or resonant doping \cite{sanchez2016nonmagnetic}. Earlier works involving TM deposition on the surface of TI reported unperturbed and gapless TSS, declaring absence of magnetic gap \cite{valla2012photoemission,wang2015robust,scholz2012tolerance}.

\begin{figure*}
	\centering
		\includegraphics[scale=0.93]{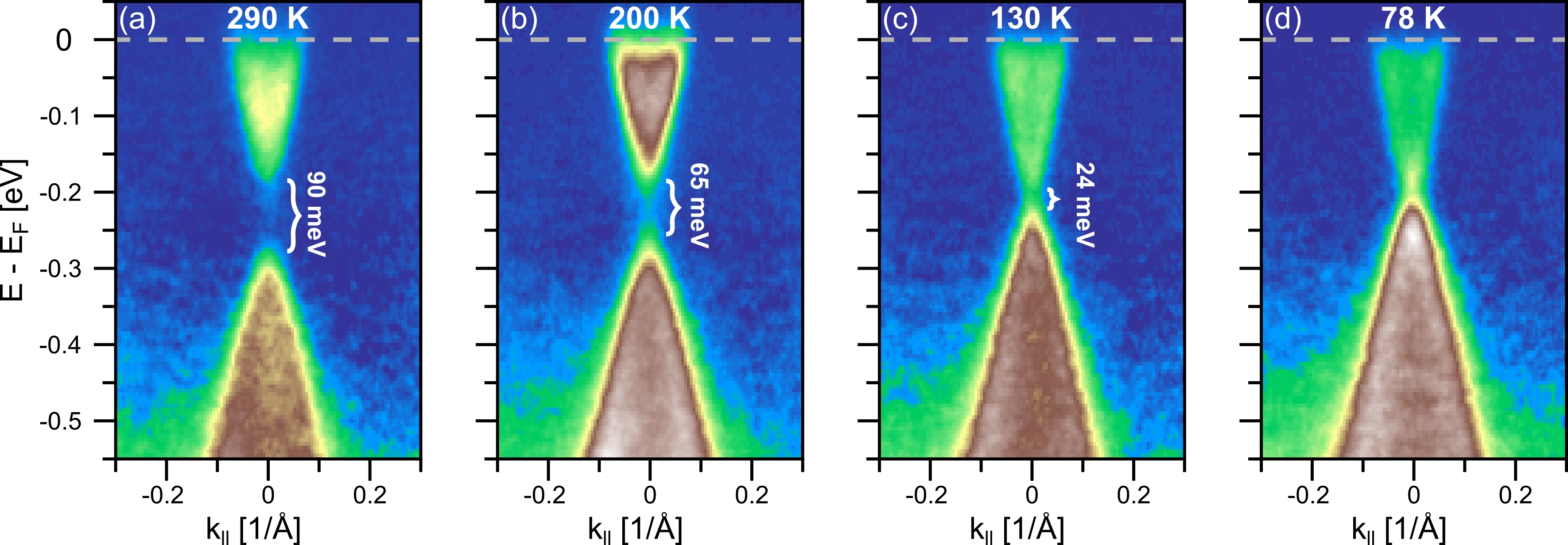}
	\caption{Temperature dependent electronic structure of Pb$_{0.75}$Sn$_{0.25}$Se with 0.2~\AA\ of Fe deposited on the surface. E($k_{||}$) spectra taken with a photon energy of 70~eV in the vicinity of the $\overline{\Gamma}$ point. The spectra are shown with a linear color scale.}
	\label{FIG:6}
\end{figure*}

\begin{figure}
	\centering
		\includegraphics[scale=0.93]{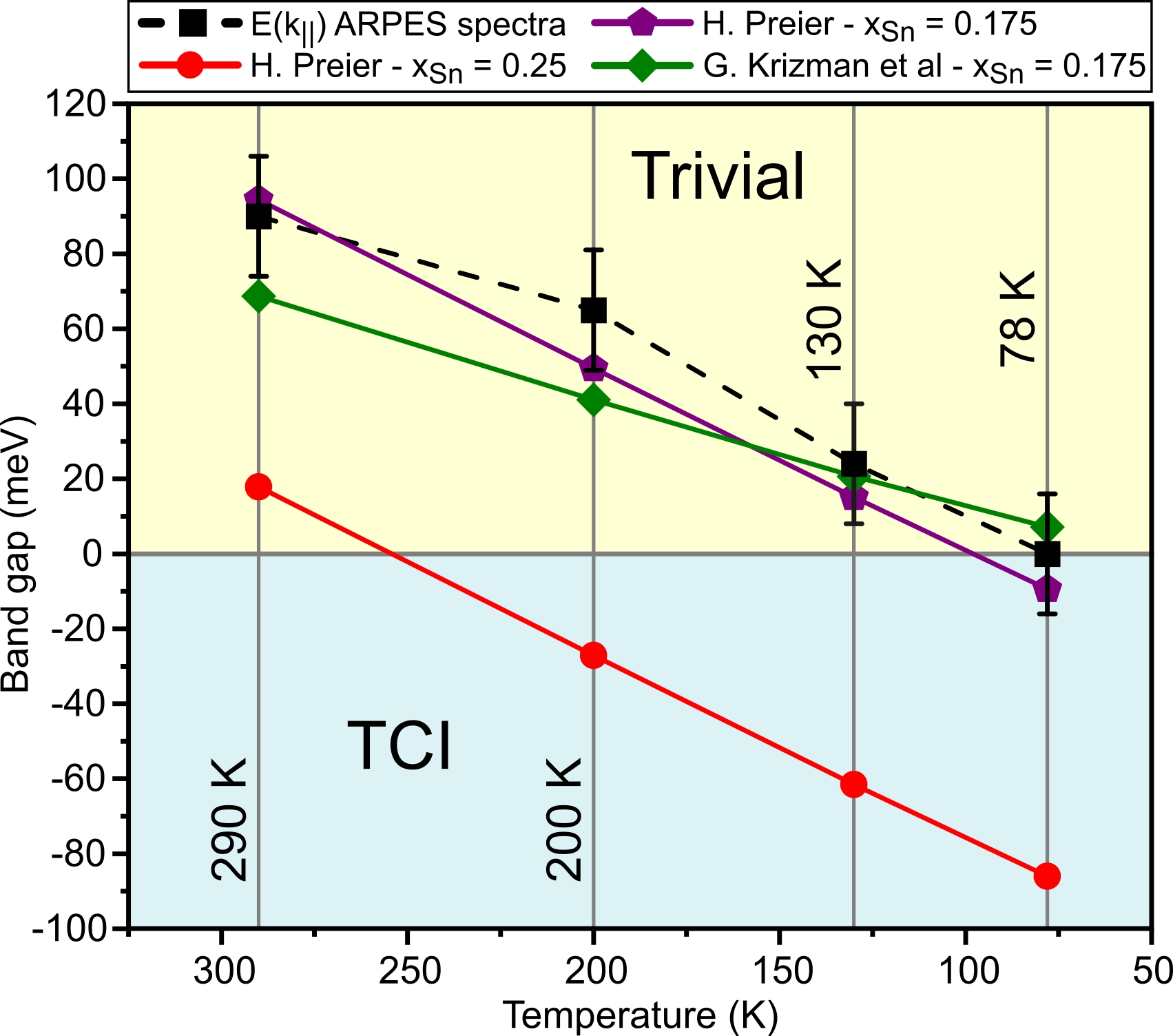}
	\caption{Temperature and Sn content dependent band gap ($E_g$) in \PbSnSe. Comparison between $E_g$ extracted from Fig. \ref{FIG:6} for x$_{Sn}$ = 0.25 with 0.2~\AA\ of Fe deposited (black dashed curve) and $E_g$ calculated following Ref. \cite{Preier1979} for x$_{Sn}$ = 0.25 (red line), for x$_{Sn}$ = 0.175 (burgundy line) and following Ref. \cite{Krizman2018} for x$_{Sn}$ = 0.175 (green line)}
	\label{FIG:7}
\end{figure}

Therefore, to investigate the origin of gap opening on TCI surface after TM deposition, temperature dependent ARPES measurements were carried out. Fig. \ref{FIG:6} (a) - (d) show evolution of E($k_{||}$) spectra with decreasing temperature for Pb$_{0.75}$Sn$_{0.25}$Se sample with 0.2~\AA\ of Fe deposited on its surface. For the epilayer of such composition, topological transition point at which band gap closing occurs is at 254.5~K according to the semi-empirical Preier's relation \cite{Preier1979}. However, we observe a band gap of 65 meV at 200 K (Fig. \ref{FIG:6} (b)). The gap reduces further with a temperature decrease and reaches 24 meV at 130~K (Fig. \ref{FIG:6} (c)). Cooling the sample down to the temperature of liquid nitrogen value results in a completely closed gap (Fig. \ref{FIG:6} (d)). Thus, the magnetic gap scenario can be excluded due to the absence of a magnetic transition point below which the band gap should remain constant. The resonance doping scenario also can be ruled out, since, for this case, the observed gap should be temperature independent \cite{sanchez2016nonmagnetic}. One should note, that the observation of the band gap induced by TM on the surface of TI is rather unusual since no gap opening was reported so far for the deposition of TM on the surface of bismuth chalcogenide-based topological compounds \cite{valla2012photoemission,wang2015robust,scholz2012tolerance}. In this respect, \PbSnSe\ TCI is quite unique. Typically, the band gap of these compounds is temperature and composition dependent. Therefore, the most probable explanation is related to a change of composition-dependent band inversion point with TM deposition. Fig. \ref{FIG:7} represents temperature dependence of $E_g$ of \PbSnSe\ TCI with x$_{Sn}$ = 0.25 and 0.175 calculated according to Preier's relation\cite{Preier1979} and more recent measurements of Ref. \cite{Krizman2018}, as well as our data extracted from Fig. \ref{FIG:6}. As one can see, our temperature dependence of the band gap reproduces the one calculated for x$_{Sn}$ = 0.175. Indeed, since TMs used in this work are relatively light elements, a decrease in spin-orbit strength is expected, hence one can consider TM deposition as a way of changing surface composition toward a trivial side. Nevertheless, we believe that more extensive experiments similar to those as in ref.~\cite{wojek2015direct} involving samples with different compositions and different amount of TM must be conducted in the future to confirm unambiguously Sn-composition dependent shift of the band inversion point with TM deposition.

\section{Conclusions}

To sum up, we investigated the electronic and spin structure of high quality TCI \PbSnSe\ (111) epilayers by means of conventional and spin-resolved ARPES. The samples that were successfully produced using the MBE method were then transferred to the synchrotron facility inside the UHV suitcase which preserved their pristine surfaces. We showed that helical spin-polarization of the surface states exists not only for topological samples but also for samples of trivial composition. The magnitude of the spin-polarization is of the same order for both types of samples. Moreover, we revealed that TM deposition on the sample surface opens the band gap, which forces TCI-NI transition. An additional point to emphasize is that we demonstrated modification of the band structure at the interface between TM and TCI. This should be taken into account in the real spintronic devices combining TCI and magnetic material based on TM.

\printcredits

\section*{Declaration of Competing Interest}
The authors declare that they have no known competing financial interests or personal relationships that could have appeared to influence the work reported in this paper.

\section*{Acknowledgement}
The work is supported by the Foundation for Polish Science through the IRA Programme co-financed by EU within SG OP (Grant No. MAB/2017/1). The authors thank NSRC SOLARIS for beamtime allocation and T. Dietl with T. Story as well as G. Springholz for valuable discussions.

\bibliographystyle{model1-num-names}

\end{sloppypar}
\end{document}